\documentclass[aip,jcp,amsmath,amssymb,reprint]{revtex4-1}
\usepackage[utf8]{inputenc} 
\usepackage[T1]{fontenc}       
\usepackage{amssymb}
\usepackage{amsmath}
\usepackage{comment} 

\usepackage{float} 
\usepackage{graphicx} 
\usepackage{color}

\usepackage{booktabs} 
\usepackage{array} 
\usepackage{paralist} 
\usepackage{verbatim} 
\usepackage{subfig} 
\usepackage{hyperref}
\usepackage[version=3]{mhchem}
\usepackage{feynmp}
\usepackage{caption}
\usepackage{ulem}

\makeatletter
\newcommand*{\rom}[1]{\expandafter\@slowromancap\romannumeral #1@}
\makeatother

\begin{document}
\title{How to correct Ehrenfest nonadiabatic dynamics in open quantum systems: Ehrenfest plus random force (E$+$$\sigma$) dynamics}

\author{Jingqi Chen}
\affiliation{Department of Chemistry, School of Science, Westlake University, Hangzhou, Zhejiang 310024, China}
\affiliation{Institute of Natural Sciences, Westlake Institute for Advanced Study, Hangzhou, Zhejiang 310024, China}

\author{Joonho Lee}
\email{joonholee@g.harvard.edu}
\affiliation{Department of Chemistry and Chemical Biology, Harvard University, Cambridge, MA 02138, USA}

\author{Wenjie Dou}
\email{douwenjie@westlake.edu.cn} 
\affiliation{Department of Chemistry, School of Science, Westlake University, Hangzhou, Zhejiang 310024, China}
\affiliation{Department of Physics, School of Science, Westlake University, Hangzhou 310024 Zhejiang, China}
\affiliation{Institute of Natural Sciences, Westlake Institute for Advanced Study, Hangzhou, Zhejiang 310024, China}

\begin{abstract}
One key challenge in the study of nonadiabatic dynamics in open quantum systems is to balance computational efficiency and accuracy. 
Although Ehrenfest dynamics (ED) is computationally efficient and well-suited for large complex systems, ED often yields inaccurate results. 
To address these limitations, we improve the accuracy of the traditional ED by adding a random force (E$+$$\sigma$). 
In this work, the construction of random forces is considered in Markovian and non-Markovian scenarios, and we ensure the dynamics satisfy the detailed balance in both scenarios. 
By comparing our E$+$$\sigma$ with existing methods such as the electronic friction model and surface hopping, we furthermore validate its reliability.
In addition, the E$+$$\sigma$ model still retains the high efficiency of ED and does not incur much additional computation.
We believe that this method provides an alternative to accurately describe the mixed quantum-classical dynamics in open quantum systems, particularly for large complex systems.

\end{abstract}

\maketitle

\section{Introduction}

Ehrenfest dynamics (ED), also known as mean field dynamics, are widely applied in the study of open quantum systems (OQSs) \cite{verdozzi2006classical, dundas2009current, scarlatella2021dynamical, choi2015ehrenfest, bellonzi2016assessment, vallejo2024evolution}. 
The ED framework simplifies the complex forces acting on the system by substituting them with mean-field forces. These forces are mathematically expressed as the expectation value of the derivative of the system Hamiltonian with respect to the nuclear coordinates, averaged over the electronic density matrix \cite{ehrenfest1927bemerkung}.
By approximating interactions with their mean values, ED is often computationally efficient to provides the overall dynamics when investigating complex systems with numerous interacting components or with high degrees of freedom \cite{scarlatella2021dynamical}.
Consequently, ED is an indispensable tool for theoretical modeling and guiding experimental investigations.

Although ED is efficient enough, why does it often fail to meet high accuracy requirements? More specifically, why cannot it successfully achieve detailed balance? The root of this limitation lies in its inability to properly account for statistical fluctuations between electronic and nuclear degrees of freedom, as ED relies on a mean field approximation \cite{parandekar2005mixed, bellonzi2016assessment}.
Recently, several approaches have been proposed to improve the accuracy of ED. 
One notable approach is derived using the Meyer-Miller-Stock-Thoss (MMST) transformation, which involves bosonizing the electronic degrees of freedom \cite{meyera1979classical, stock1997semiclassical}. 
This method, also known as the Poisson-Bracket Mapping Equation (PBME) \cite{kim2008quantum, kim2014improving}, often struggles in systems with strong electronic state interactions, particularly in high-dimensional or complex systems where such couplings dominate the dynamics.
Additionally, Miller and colleagues \cite{cotton2013symmetrical, cotton2013symmetrical} proposed another refinement to ED, which is known as symmetrical quasi-classical windowing (SQC).
SQC builds on the idea of symmetrically windowing both the initial and final trajectories, akin to a quasi-classical vibrational analysis, to incorporate zero-point energy effects. 
This modified method incorporates quantum effects but remains sensitive to the choice of window function, which limits its accuracy and applicability in complex systems \cite{he2024nonadiabatic,xie2018performance,weight2021ab}.

With this background in mind, we provide a new approach to address the inaccuracy of ED by incorporating a random force. 
This idea is inspired by the electronic friction (EF) model, which introduces a stochastic term to account for the coupling between a quantum system and the environment \cite{tully2000chemical, maurer2017mode, dou2015frictional,bode2012current,dou2018perspective}. 
By adding a random force, we can introduce fluctuations that help restore the detailed balance and thus improve the accuracy of the dynamics, particularly in OQSs where interactions with the environment play a significant role.
We name it as E$+$$\sigma$ is because people open use $\sigma$ as the standard deviation of a Gaussian noise.
Compared with the PBME method, the E$+$$\sigma$ method does not require an additional "Bose" process of electron freedom when dealing with complex systems, so the computational cost is lower and it is suitable for larger-scale systems or long-term dynamic simulations. 
As mentioned earlier, the window function treatment in the SQC method may introduce additional uncertainties and tend to be highly sensitive to parameter selection. In contrast, our E$+$$\sigma$ method exhibits relatively low sensitivity to parameter tuning, ensuring more consistent performance across various applications.
Moreover, our E$+$$\sigma$ framework is naturally extensible and can be further combined with other correction methods (such as PBME or SQC) to supplement its deficiencies in specific situations. This flexibility makes the E$+$$\sigma$ method easier to apply and promote in different research fields.

In this work, we consider utilizing the E$+$$\sigma$ method to describe a molecule moving close to a fermionic bath (metal).
The simplest case is that a single impurity level is coupled to a single phonon as well as one or two fermionic baths, which is known as the Anderson-Holstein (A-H) model. 
In the A-H model, the system-bath coupling strength ($\Gamma$) plays a pivotal role in determining the Markovity. When $\Gamma \gg \hbar \omega$ (with $\hbar \omega$ representing the nuclear frequency), random force exhibits Markovian behavior. This means that the future behavior of the random force depends solely on its current state and is not influenced by its past states, which means that the memory effects or correlations between different time steps are negligible.
Furthermore, it is worth noting that in our previous discussion, we highlighted the equivalence of the Markovian random force for ED and EF, as outlined in Ref. \!\!\citenum{dou2018perspective}.
Combined with the above, constructing a Markovian random force for ED is easy to implement.
That being said, when the strong coupling condition is not met, we must consider constructing a non-Markovian random force, which is mainly based on its memory kernel $\mathcal{K}$.

When developing a modified ED approach, it is crucial to establish benchmarks for comparisons to assess its effectiveness and accuracy. To this end, we consider two prominent methods representing different regimes of $\Gamma$: surface hopping (SH) \cite{shenvi2009dynamical, shenvi2009nonadiabatic, ouyang2015surface, dou2015surface, dou2015surface2} and the electronic friction (EF) model \cite{dou2015frictional, dou2017born, dou2018perspective}. SH, which is widely employed in the weak system-bath coupling regime, simulates the dynamics of nonadiabatic processes by probabilistically switching between different electronic states. 
However, the EF model, utilized in the strong system-bath coupling regime, models the dissipative effects of the environment on electronic motion by introducing a stochastic force. 
By selecting these two benchmark methods, we cover a spectrum of coupling strengths, allowing for a comprehensive evaluation of the proposed modified ED across different coupling regimes \cite{dou2020nonadiabatic}.

The remainder of this paper is organized as follows:
in Sec. \ref{sec:theory}, we first introduce the general method of random force construction for ED. 
We then introduce the A-H model.
In addition, we present the classical master equation (CME) and briefly introduce the CME-SH approach and the CME-EF model.
Next, we derive the ED in terms of the CME and provide a well-founded conjecture about the non-Markovian random force for the A-H model.
Furthermore, we provide a general method to generate the non-Markovian random force.
In Sec. \ref{sec:results}, we report the numerical results obtained by each method and compare them.
Finally, we conclude this paper in Sec. \ref{sec:conclusions}.

\section{Theory} \label{sec:theory}

\subsection{Ehrenfest dynamics with random force}
In the mixed quantum-classical molecular dynamics, the trajectories of nuclei ($R_{\alpha}$, $P_{\alpha}$) are governed by Newton's equations, while the evolution of the electronic density matrix ($\hat{\rho}$) is described by the von Neumann-Liouville equation
\begin{eqnarray}
\label{eq_newton}
    -m_{\alpha}{\ddot{R}_{\alpha}} = \partial_{\alpha}\hat{H}, \\
    \frac{d}{dt} \hat{\rho} = - \frac{i}{\hbar} [ \hat{H}, \hat{\rho}],
\end{eqnarray}
where $\hat{H}$ denotes the general electronic Hamiltonian, and $\alpha$ is the index for nuclear degrees of freedom (DoFs).
Eq. \ref{eq_newton} originates from the Heisenberg equation of motion, and the traditional Ehrenfest dynamics (ED) framework deviates from it because the average force in ED is $- \mathrm{Tr}_e (\partial_{\alpha} \hat{H} \hat{\rho}) $. Based on that deviation, we can thereby define the random force $\delta \hat{F}_{\alpha} = -\partial_{\alpha} \hat{H} + \mathrm{Tr}_e (\partial_{\alpha} \hat{H} \hat{\rho}) $. 
Considering the slow nuclear motion, we can further make the adiabatic approximation to substitute $\hat{\rho}$ with the steady-state density matrix  $\hat{\rho}_{ss}$,
\begin{equation}
\label{eq:df_rf}
    \delta \hat{F}_{\alpha} = -\partial_{\alpha} \hat{H} + \mathrm{Tr}_e (\partial_{\alpha} \hat{H} \hat{\rho}_{ss}).
\end{equation}
Now we can subsequently rewrite Eq. \ref{eq_newton} as
\begin{equation}
\label{eq_ehren}
    -m_{\alpha}{\ddot{R}_{\alpha}} = \mathrm{Tr}_e (\partial_{\alpha} \hat{H} \hat{\rho}) - \delta \hat{F}_{\alpha},
\end{equation}
so we understand that the inaccuracy of the traditional ED stems from the absence of an random force term, with errors accumulating over time. 
In the above equation, everything is quantum mechanics. In the ED, the random force is neglected, such that one makes the classical approximation to the nuclear motion. Similar to ED, we can replace the operators by their classical corresponds: Both the nuclear momentum and the random force are replaced by the classical ones. Such a statement is commonly used in mixed quantum-classical dynamics \cite{bode2011scattering, dou2018perspective,lu2012current, dou2017born}.
We mainly focus on how to construct this random force to correct the traditional ED, thus implementing our novel E$+$$\sigma$ method.
The operator $\delta \hat{F}_{\alpha}$ can potentially be substituted with a random number $\xi_{\alpha}$. 
To achieve this, the relevant correlation function of the random force must be calculated. 
In the context of the suggested mixed quantum-classical dynamics approach, we postulate that the fitting correlation function can be represented by a symmetrized product
\begin{eqnarray}
\label{eq_correlation}
    \begin{aligned}
        D_{\alpha \nu}^S (t, t^{\prime}) = \frac{1}{2} \mathrm{Tr}_e 
        \bigl( \delta \hat{F}_{\alpha} (t) (\delta \hat{F}_{\nu} (t^{\prime}) \hat{\rho}_{ss} + \hat{\rho}_{ss} \delta \hat{F}_{\nu} (t^{\prime})) \bigr) .
    \end{aligned}
\end{eqnarray}
To practically implement the evolution of ED, random force $\delta \hat{F}_{\alpha}$ can be simulated by extending a random number $\xi_{\alpha}$, ensuring that its statistical properties satisfy the following conditions
\begin{eqnarray}
    \begin{aligned}
    \left\langle{\xi_{\alpha}(t)}\right\rangle &= 0,\\
    \frac{1}{2}\left\langle{\xi_{\alpha}(t)\xi_{\nu}(t^{\prime}) + \xi_{\nu}(t)\xi_{\alpha}(t^{\prime})}\right\rangle &= D_{\alpha \nu}^S (t-t^{\prime}).        
    \end{aligned}
\end{eqnarray}
So far, we have presented a general picture of our E$+$$\sigma$ method. If we further apply the Markovian approximation, when the decay of the memory kernel is faster than the nuclear motion: $D_{\alpha \nu}^S (t, t^{\prime}) \rightarrow  2 [\overline{D}^S_{\alpha \nu}]_M \delta (t-t^{\prime}) $, so Eq. \ref{eq_correlation} can be simplified as 
\begin{equation}
[\overline{D}^S_{\alpha \nu}]_M = \frac{1}{2} \int_{0}^{\infty} \mathrm{Tr}_e \bigl( \delta \hat{F}_{\alpha} (t) (\delta \hat{F}_{\nu} (0) \hat{\rho}_{ss} + \hat{\rho}_{ss} \delta \hat{F}_{\nu} (0) \bigr)dt .
\end{equation}
The random force is introduced to restore the detailed balance that is inherently broken in ED. By simulating quantum fluctuations between the nuclei and electrons, the random force supplements the missing stochasticity in the mean field approximation. 

\subsection{Anderson-Holstein model}
To illustrate the corrective approach for the ED method, we utilize the Anderson-Holstein (A-H) model \cite{anderson1961localized, holstein1959studies} as an example. 
The A-H model offers a comprehensive framework for studying a molecule featuring a single electronic level, engaged in interactions with both an electron continuum and a collection of nuclear degrees of freedom (DoFs).
The total Hamiltonian, denoted as $\hat H_{\text{tot}}$, can be divided into three parts: the system $(\hat{H_{\text{s}}})$, the continuous levels of electrons to be bath $(\hat{H_{\text{b}}})$, and the coupling between them $(\hat{H_{\text{c}}})$,
\begin{eqnarray}
\label{e5}
\begin{aligned}
&\hat H_{\text{tot}} = \hat{H_{\text{s}}} + \hat{H_{\text{b}}} + \hat{H_{\text{c}}}, \\
&\hat{H_{\text{s}}} = h(R)\hat d^{\dagger}\hat d + U_0(R) + \sum_{\alpha}\frac{P_{\alpha}^{2}}{2m_{\alpha}},\\
&\hat{H_{\text{b}}} = \sum_{k}(\epsilon_{k}-\mu)\hat c^{\dagger}_{k}\hat c_{k}, \\
&\hat{H_{\text{c}}} = \sum_{k}V_{k}(\hat c^{\dagger}_{k}\hat d + \hat c_{k}\hat d^{\dagger}).
\end{aligned}
\end{eqnarray}
Here, $\hat d^{\dagger} \hat (d)$, $\hat c^{\dagger}_k \hat(c_k)$ are creation (annihilation) operators on impurity electrons and metal electrons, respectively. 
$R$ and $P$ are the nuclear position and momenta, and we use $\alpha$ as its DoF.
$U_0(R)$ is the neutral nuclear potential for a molecule when $\hat d^{\dagger}\hat d = 0$; 
when it is charged, $\hat d^{\dagger}\hat d = 1$ and $U_1(R) = U_0(R) + h(R)$.
$V_k(R)$ represents the coupling between the molecular orbital $d$ and the metallic orbital $k$, which generally depends on $R$. 
$\epsilon_{k}$ is the energy level of the bath, which has Fermi level $\mu$, and $V_k$ is the coupling 
between the impurity and the bath. 
The interaction between the impurity and bath determines the hybridization function $\Gamma$,
\begin{eqnarray}
\label{e6}
\Gamma(\epsilon) = 2\pi\sum_{k}|V_k|^2\delta(\epsilon-\epsilon_k).
\end{eqnarray}
To further simplify the interaction, we use the wide-band approximation to treat $\Gamma$ as a constant, so it no longer depends on $\epsilon$. 
In addition, we must mention the regimes of the system-bath coupling. 
The typical timescale for electronic motion is characterized by $\hbar/\Gamma$ while the typical timescale for 
the nuclear motion can be quantified as $1/\omega_{\alpha}$, 
where we define the nuclear frequency as equal to $\sqrt{\partial^{2}_{\alpha}U_0/M_{\alpha}}$.
When we talk about the strong molecule-metal interaction regime, here we can briefly consider it as $\Gamma  \gg  \hbar\omega_{\alpha}$;
similarly, $\Gamma  \ll  \hbar\omega_{\alpha}$ indicates a weak molecule-metal interaction.

\subsection{Classical master equation/surface hopping} 
The classical master equation (CME) is based on a perturbative expansion in the metal-molecule coupling and is accurate when the molecular level broadening due to metal-molecule charge-transfer interaction can be neglected, i.e., when $kT \gg \Gamma$ and $kT \gg \hbar \omega_{\alpha}$.
Under this high-temperature limitation, we can provide a CME for the coupled electron-nuclear dynamics of the molecule described by the AH model.
See Ref. [\citenum{elste2008current}] for the detail of CME.
We firstly define our system Hamiltonian: $H_{\zeta} = U_{\lambda} + \sum_{\alpha}\frac{P^2_{\alpha}}{2m_{\alpha}}, \lambda = 0,1 $. 
Then we have
\begin{eqnarray}
\label{e7}
\begin{aligned}
    \frac{\partial \rho_0(R,P,t)}{\partial t} =& \{ H_0(R,P), \rho_0(R,P)\}\\& - \frac{\Gamma}{\hbar}f(h)\rho_0 + \frac{\Gamma}{\hbar}(1-f(h))\rho_1, \\  
    \frac{\partial \rho_1(R,P,t)}{\partial t} =& \{ H_1(R,P), \rho_1(R,P)\} \\&+ \frac{\Gamma}{\hbar}f(h)\rho_0 - \frac{\Gamma}{\hbar}(1-f(h))\rho_1, 
\end{aligned}
\end{eqnarray}
where $\rho_0(R,P)$ and $\rho_1(R,P)$ are the probability densities in the phase space for the neutral ($\hat d^{\dagger}\hat d = 0$) 
and charged ($\hat d^{\dagger}\hat d = 1$) molecules, and $\left\{ \cdot,\cdot \right\}$ represents the Poisson bracket.
These densities follow classical motion on potential energy surfaces $U_0(R)$ and $U_1(R)$ correspondingly, 
plus exchange between the two densities. 
The exchange rate from $\rho_0(R,P)$ to $\rho_1(R,P)$ is $\frac{\Gamma}{\hbar}f(h)$, and the exchange rate from $\rho_1(R,P)$ to $\rho_0(R,P)$ is $\frac{\Gamma}{\hbar}(1-f(h))$. 
Additionally, $f(h)$ denotes the Fermi-Dirac distribution function,
$f(h) = (e^{(h-\mu)/kT}+1)^{-1}$, where $kT$ represents the temperature.

We now introduce surface hopping (SH) algorithms to solve the CME.
If we use a group of trajectories to indicate the density probability, each trajectory follows a classical Newtonian motion on one of the two PESs:
\begin{eqnarray}
\begin{aligned}
    \dot{R}_{\alpha}=\frac{P_{\alpha}}{M_{\alpha}}, 
    \dot{P}_{\alpha}=-\frac{\partial U_{\zeta}}{\partial R_{\alpha}},
\end{aligned}
\end{eqnarray}
where the index $\zeta = 0$ or $1$ indicates the potential energy surface. The rate of switching from $U_0$ to $U_1$ is  $\frac{\Gamma}{\hbar}f(h)$, and the rate of switching from $U_1$ to $U_0$ is $\frac{\Gamma}{\hbar}(1-f(h))$.
See Ref. [\citenum{dou2015surface}] for details on CME-SH. In this study, the CME-SH method is used for benchmark in the regime that $kT \gg \Gamma$ and $kT \gg \hbar \omega_{\alpha}$.

\subsection{Fokker-Planck equation/electronic friction}\label{subsec:friction}
In the adiabatic limit, we can trace out all of the electronic DoFs, and obtain the general Langevin equation in the Markovian limit: 
\begin{eqnarray}
\label{e10}
    -m_{\alpha}\ddot{R}_{\alpha} = -\overline F_{\alpha}+ \sum_{\nu} \gamma_{\alpha\nu}\dot{R}_{\nu}-\xi_{\alpha}.
\end{eqnarray}
Under the CME limitations $kT \gg \hbar\omega_{\alpha}$, we can use CME to derive the Fokker-Planck equation,
\begin{eqnarray}
\begin{aligned}
   \frac{\partial \mathcal{A}}{\partial t} =& - \sum_{\alpha}\frac{P_{\alpha}}{M_{\alpha}} - \sum_{\alpha} \overline F_{\alpha}\frac{\partial \mathcal{A}}{\partial P_{\alpha}}  \\
   &+ \sum_{\alpha\nu}\gamma_{\alpha\nu} \frac{\partial}{\partial P_{\nu}}(\frac{P_{\alpha}}{M_{\alpha}} \mathcal{A})+ 
   \sum_{\alpha\nu}D_{\alpha\nu} \frac{\partial^{2} \mathcal{A}}{\partial R_{\alpha}\partial R_{\nu}},
\end{aligned}   
\end{eqnarray}
and finally arrive at the CME-electronic friction (EF) model. Here, the mean force $\overline F_{\mu}$,
friction tensor $\gamma_{\alpha\nu}$, and correlation function of the Markovian random force $[\overline{D}_{\alpha\nu}^{S}]_M$ are expressed as
\begin{eqnarray}
\label{eq:friction_model}
\begin{aligned}
    \overline F_{\alpha} =& - \frac{\partial U_0}{\partial R_{\alpha}}-\frac{\partial h}{\partial R_{\alpha}}f(h), \\  
    \gamma_{\alpha\nu} =& -\frac{\hbar}{\Gamma}\frac{\partial f(h)}{\partial R_{\nu}}\frac{\partial h}{\partial R_{\alpha}}, \\
    [\overline{D}_{\alpha\nu}^{S}]_M =& \frac{\hbar}{\Gamma}f(h)(1-f(h))\frac{\partial h}{\partial R_{\alpha}}\frac{\partial h}{\partial R_{\nu}}.
\end{aligned}
\end{eqnarray}
Besides, $\xi_{\alpha}$ denotes the random force, which is assumed to be a Gaussian variable with a norm $\sigma_{\alpha} = \sqrt{2 \hbar  [\overline{D}_{\alpha\nu}^{S}]_M / \omega_{\alpha} dt}$.
See Ref. \citenum{dou2015frictional} for the full derivation of the EF model in terms of the CME.

\subsection{E$+$$\sigma$ for AH model}
\label{sec:rf1}
In Sec. \ref{sec:rf1}, we will discuss how to specifically model the E$+$$\sigma$ dynamics for the A-H model, which has only one energy level.
We firstly rewrite Eq. \ref{eq_ehren}, in the following format:
\begin{eqnarray}
\label{e13}
    m_{\alpha}{\ddot{R}_{\alpha}} = -(\rho_0(R,P) \frac{\partial U_0}{\partial R_{\alpha}} + \rho_1(R,P) \frac{\partial U_1}{\partial R_{\alpha}}) + \xi_{\alpha},
\end{eqnarray} 
where $\rho_0(t)$ and $\rho_1(t)$ can be obtained by evolving the CME. Additionally, we stress the importance of ensuring that the added random force for correcting the ED strictly conforms to the random force in the EF model. The calculation of the random number $\xi_{\alpha}$ under Markovian approximation is provided in Sec. \ref{subsec:friction}. 
Beyond the Markovian limit, we still need to determine how to compute the non-Markovian random force in general.

We consider generating a Markovian random number $[{\xi}_{\alpha}]_M$ at each time step according to the correlation function of the Markovian random force $[\overline{D}^S_{\alpha \nu}]_M$, which allows for stochastic interactions between the system and the bath. Meanwhile, introducing non-Markovian processes accounts for memory effects within the system, and the memory kernel $ \mathcal{K} (t)= [\overline{D}^S_{\alpha \nu}]_M / {(2 \tau_e)} e^{- t / \tau_e}$ ($\tau_e $ is the timescale for electronic motion). In our case,  $\tau_e = \hbar / \Gamma$, thus we have
\begin{equation}
\label{eq_kernel}
    \mathcal{K}(t) = \frac{  [\overline{D}^S_{\alpha \nu}]_M \Gamma}{2 \hbar}e^{- \Gamma t/ \hbar } .
\end{equation}
In the spectrum of the memory kernel, $\mathcal{K}(t)$ corresponds to  colored noise, whereas $[\overline{D}^S_{\alpha \nu}]_M$ is responsible for white noise \cite{ceriotti2009langevin}.  
Thus, we must construct a stochastic differential equation (SDE) to evolve the non-Markovian random number $[{\xi}_{\alpha}]_N$ that satisfies the memory kernel shown in Eq. \ref{eq_kernel}. 
This dual approach captures both short- and long-term influences on the system dynamics, leading to a more accurate simulation of its behavior.
For those purposes, we construct a non-Markovian random force as follows:
\begin{eqnarray}
\label{eq:generate_rf}
 [\dot{\xi}_{\alpha}]_N = -\Gamma ([{\xi}_{\alpha}]_N - [{\xi}_{\alpha}]_M).
\end{eqnarray}
We ensure that the integral of Eq. \ref{eq:generate_rf} can be returned to Eq. \ref{eq_kernel}.
At each time step, Eq. \ref{eq:generate_rf} is first solved to update the random force, followed by the propagation of the ED using Eq. \ref{e13}.
So far, we have understood how to specifically construct a non-Markovian random force for the AH model under the wide-band approximation, but it is not general. Next, we will provide a more general method to construct the random force in Sec. \ref{sec:rf2}.

\subsection{General E$+$$\sigma$: numerically calculate random force}
\label{sec:rf2}
\begin{figure}
  \centering
  \includegraphics[width=0.45\textwidth]{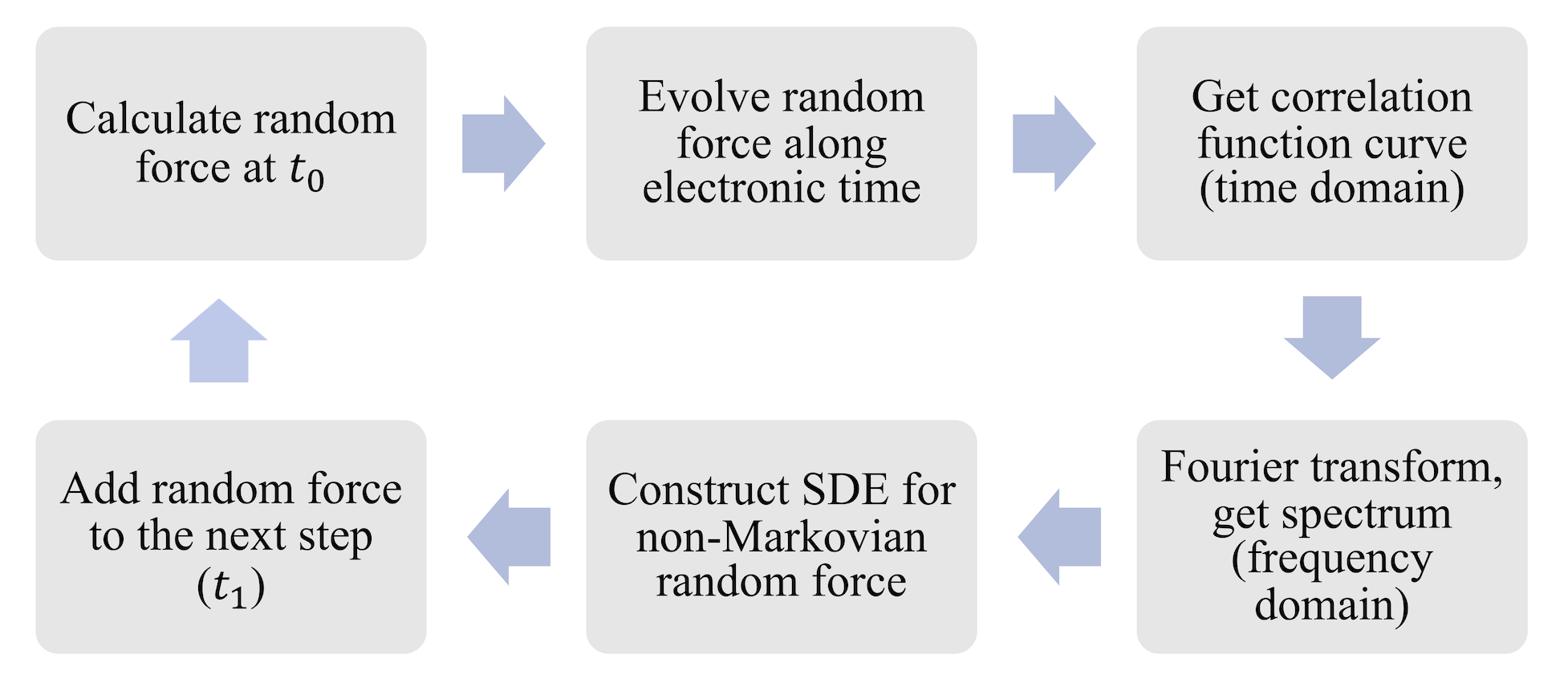}
  \caption{A flowchart which explains how to generally generate a non-Markovian random force.}
  \label{fig:Flowchart}
\end{figure}

If the correlation function is simple and its analytical form is known, in such cases, as has been demonstrated in Sec. \ref{sec:rf1}, we can directly construct the evolution equation for the non-Markovian random force analytically. The method presented in Sec. \ref{sec:rf2} here is intended for more complex systems, where the mathematical form of the correlation function is not straightforward.
Here, we still utilize the AH model but suppose that we do not know the specific math form of the memory kernel $\mathcal{K} (t)$ (Eq. \ref{eq_kernel}) and how to calculate
$[\overline{D}^S_{\alpha \nu}]_M$ by EF model (Eq. \ref{eq:friction_model}) as well.
We will provide a more general method to get the SDE (Eq. \ref{eq:generate_rf})  in Sec. \ref{sec:rf2}. The SDE is constructed according to the spectrum of the random force correlation function in the frequency domain, allowing for the generation of a Markovian random force at each time step and propagation of the random force according to the differential equation for the non-Markovian random force. 
Fig. \ref{fig:Flowchart} is the flowchart to show the procedures.
Next, we will describe each step in detail in combination with our A-H model.

The first step is to extract the random force. 
Based on the definition of the random force presented in Eq. \ref{eq:df_rf}, and under the classical approximation of the system Hamiltonian, we have
\begin{eqnarray}
\begin{aligned}
[\delta F_{\alpha}]_0 &= -\frac{\partial H_0}{\partial R_{\alpha}} + \text{Tr}_e\left(\frac{\partial H}{\partial R_{\alpha}} \rho_{ss}\right), \\
[\delta F_{\alpha}]_1 &= -\frac{\partial H_1}{\partial R_{\alpha}} + \text{Tr}_e\left(\frac{\partial H}{\partial R_{\alpha}} \rho_{ss}\right).
\end{aligned}
\end{eqnarray}
Then we try to numerically calculate the correlation function. According to the definition of the quantum correlation function, 
\begin{eqnarray}
\begin{aligned}
<[\delta F_{\alpha}](t)[\delta F_{\alpha}](t + \tau)>  = 
&\sum_{\zeta} e^{-i\hat{H}_{\text{tot}} \tau} [\delta F_{\alpha}]_{\zeta}(t)[\rho_{ss}]_{\zeta}(t)  \\
&\times e^{i\hat{H}_{\text{tot}} \tau}  [\delta F_{\alpha}]_{\zeta}(t).
\end{aligned}
\end{eqnarray}
$\zeta$ can be $0$ or $1$ to represent different surfaces. Here we cannot directly use the total Hamiltonian $\hat{H}_{tot}$ to propagate the random force from time $t$ to time $t + \tau$, so we consider using the CME to evolve $\delta F_{\alpha}(t)\rho_{ss}(t)$ together. Since we know that $[\rho_{ss}]_0 =1-f(h)$, $[\rho_{ss}]_1 = f(h)$, then we get
\begin{eqnarray}
\begin{aligned}
\frac{\partial([\delta F_{\alpha}]_0 [\rho_{ss}]_0 )}{\partial \tau} =   &\frac{\Gamma}{\hbar}f(h)(1-f(h))  \\
&\times  (-[\delta F_{\alpha}]_0 + [\delta F_{\alpha}]_1), \\
\frac{\partial([\delta F_{\alpha}]_1 [\rho_{ss}]_1)}{\partial \tau} =   &\frac{\Gamma}{\hbar}f(h)(1-f(h))  \\
&\times ([\delta F_{\alpha}]_0 - [\delta F_{\alpha}]_1).
\end{aligned}
\end{eqnarray}
Although we select CME to compute the correlation function in this work, the choice of approach is not unique. 
Although we utilized the CME approach to compute the correlation function in this work, the choice of approach for calculating the correlation function is not unique. For instance, one could calculate the correlation functions from IESH or BCME, which will incorporate broadening effects that are missing in CME.  
As a side note, we must emphasize that the calculation of the random force correlation function and the propagation of dynamics are not parallel. Instead, the correlation function must be pre-computed, which is used to generate random force in the final equation of motion.

For our case that the system-bath coupling is a constant under the wide-band approximation, the correlation function exhibits no oscillations and decays exponentially in the time domain \cite{ceriotti2009langevin}. 
Due to no oscillation, there is only one peak at the zero point. Thus, the Fourier transform result is in a Cauchy distribution form (as is shown in the next section)
\begin{eqnarray}
\label{eq:kernel_w}
\begin{aligned}
\mathcal{K} (\omega_s)
= \frac{2 [\overline{D}^{S, \prime}_{\alpha \nu}]_M [\Gamma^{\prime}]^2}{\omega^2_s+[\Gamma^{\prime}]^2}.
\end{aligned}
\end{eqnarray}
We can fit the parameters $ [\overline{D}^{S, \prime}_{\alpha \nu}]_M$  and $\Gamma^{\prime}$ by the obtained power spectrum data as we have known its mathematical form. In addition, Eq. \ref{eq:kernel_w} can be Fourier transformed back to Eq. \ref{eq_kernel}, thereby get the SDE shown in Eq. \ref{eq:generate_rf} to evolve the non-Markovian random force.

However, if we consider more general and complex situations, which means the correlation function exhibits oscillations, then obtaining the expression directly becomes hard. We must resort to transforming to the frequency domain for further analysis. Here, we provide an example to show how to gain information after Fourier transformation.
For a more general case, where the frequency domain spectrum has a series of peaks at $\omega_0$, $\omega_1$,..., $\omega_n$, then the distribution function becomes a series of peaks \cite{wang2021light, philbin2022chemical}
\begin{eqnarray}
\begin{aligned}
\mathcal{K} (\omega_s)
= \frac{2 [\overline{D}^S_{\alpha \nu}]_M \Gamma\omega_n^3}{(\omega^2_s-\omega_n^2)^2+(\omega_s\Gamma)^2}.
\end{aligned}
\end{eqnarray}
We also need to extract the necessary parameters from the power spectrum through fitting, and then derive the SDE for the random force.
When we know the method to generate the random force at time $t$, we can add it back to the equation of motion at the next step $t+dt$. 
At the end of this section, we should point out that if the dynamics results demonstrate convergence, the update interval of $[\overline{D}^S_{\alpha \nu}]_M$ can be extended to enhance computational efficiency. This adjustment will be explored further in Sec. \ref{sec:compare}.

\section{Numerical results and discussion} \label{sec:results}

In this section, we investigate a single nuclear mode described by $U_0 = \frac{1}{2}m\omega^2x^2$ and $h(x) = E_d+gx\sqrt{2m\omega/\hbar}$ ($g$ represents the electron-phonon coupling strength). 
\begin{figure}[ht]
\centering
\includegraphics[width = 8cm]{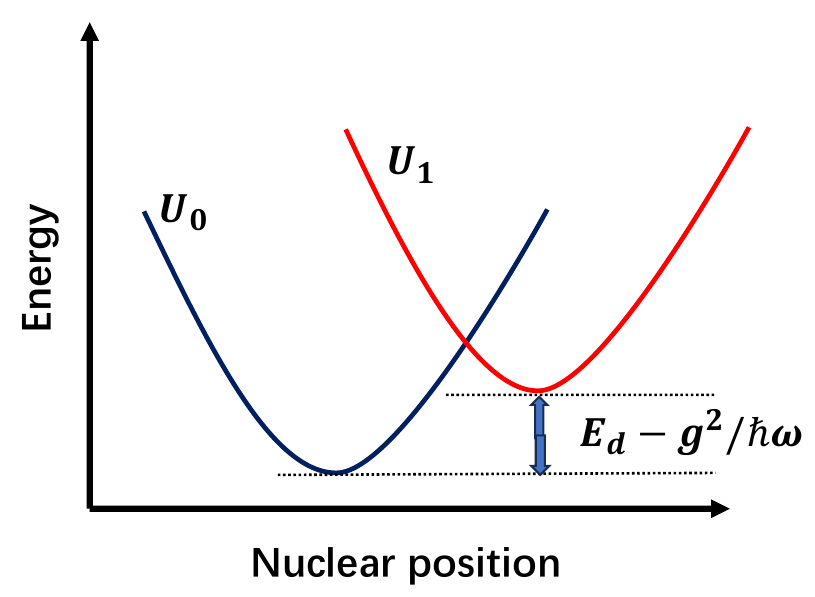}
\caption{Potential energy surfaces for the two charge states $U_0$ and $U_1$. The renormalized energy between the two nuclear modes is directly related to $E_d$. Specifically, when $E_d = \frac{g^2}{\hbar \omega}$, the renormalized energy is zero.}
\label{fig:sch}
\end{figure}

We employ the 4th-order Runge-Kutta method to numerically integrate the differential equations and utilize $50,000$ trajectories for each simulation. We compare five different dynamics methods: traditional Ehrenfest dynamics (ED), electronic friction-Langevin dynamics (EF-LD), Ehrenfest dynamics with Markovian random force (M-ED), Ehrenfest dynamics with non-Markovian random force (NM-ED), and surface hopping (SH). By the way, we should point out that the NM-ED we used in Sec. \ref{sec:case1} and Sec. \ref{sec:case2} are mentioned in Sec. \ref{sec:rf1}, in which we analytically obtain $ [\overline{D}^{S}_{\alpha \nu}]_M$. Subsequently, in Sec. \ref{sec:compare}, we will compare this method with the more general one mentioned in Sec. \ref{sec:rf2}. 

\subsection{Case 1: one electronic bath}
\label{sec:case1}
\begin{figure}[ht]
\centering
\includegraphics[width = 8cm]{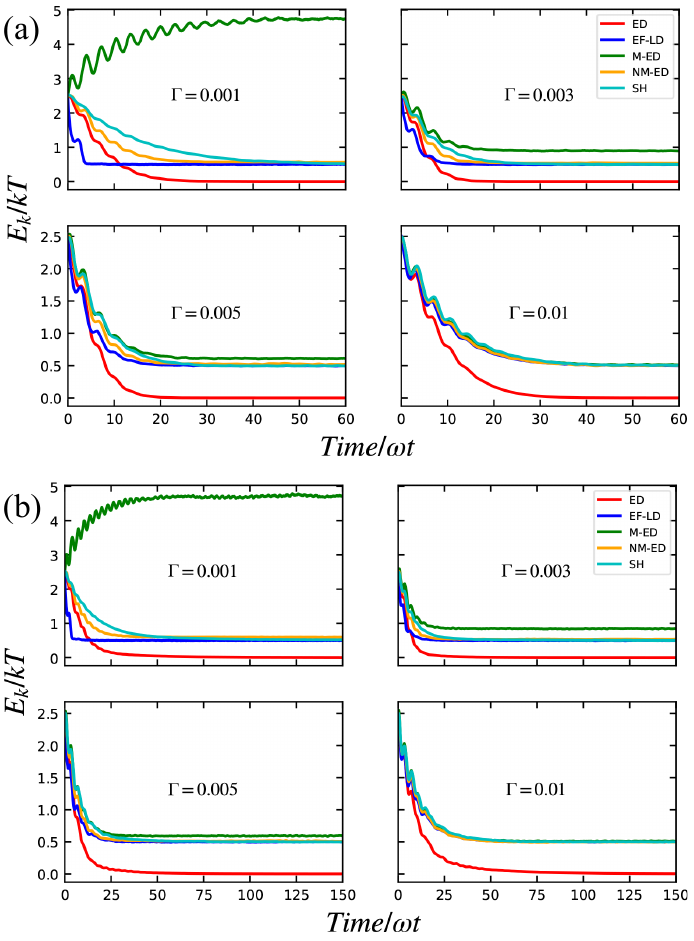}
\caption{Phonon relaxation with different system-bath coupling strength $\Gamma$. Here, we plot the average kinetic energy via five different approaches as a function of time,
which are traditional Ehrenfest dynamics (ED), electronic friction-Langevin dynamics (EF-LD), Ehrenfest dynamics with a Markovian random force (M-ED), Ehrenfest dynamics with non-Markovian random force (NM-ED), and surface hopping (SH), respectively.
Parameters: $\hbar\omega = 0.003, kT = 0.05, g = 0.02$, $\ E_d = g^2/\hbar\omega$ in (a) while $\ E_d = g^2/2\hbar\omega$ in (b). 
We prepare initial states satisfying a Boltzmann distribution with a temperature of $5 kT$. }
\label{fig:eqek}
\end{figure}

\begin{figure}[ht]
\centering
\includegraphics[width = 8cm]{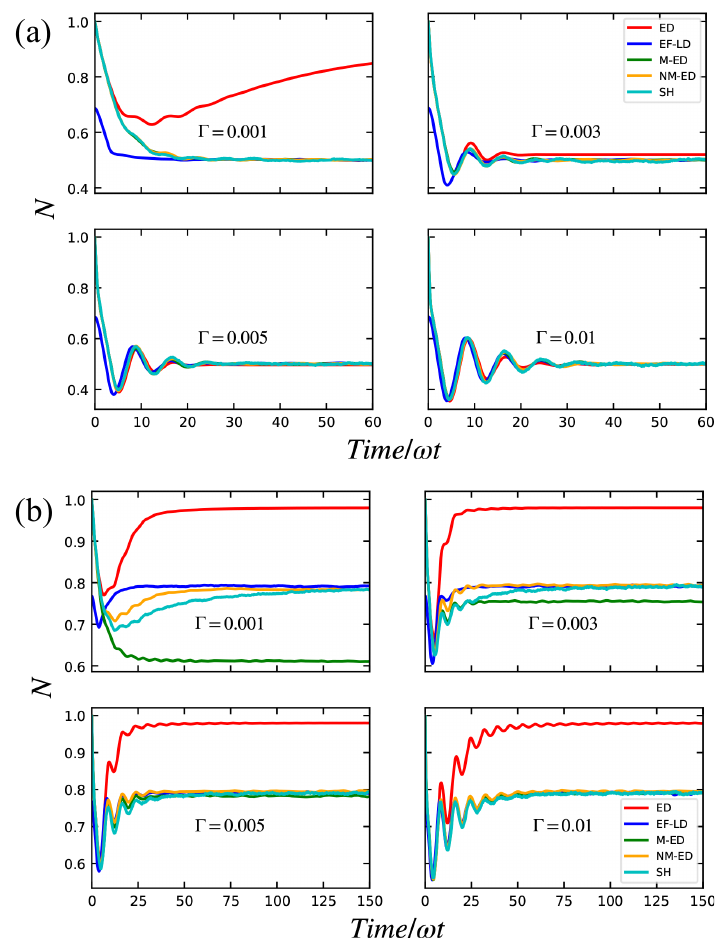}
\caption{Electronic population in the impurity level by five different trajectory-based methods. Parameters: $\hbar\omega = 0.003, kT = 0.05, g = 0.02$, $\ E_d = g^2/\hbar\omega$ in (a) while $\ E_d = g^2/2\hbar\omega$ in (b).  }
\label{fig:eqpo}
\end{figure}

We first present the dynamical results for a scenario involving only one bath surface near the molecule. 
Fig.~\ref{fig:eqek} depicts the evolution of the average kinetic energy over time. Notably, the average kinetic energy simulated by ED converges to zero regardless of the system-bath coupling strength $\Gamma$, indicating that the dissipation properties are overlooked in traditional ED. 
Additionally, EF-LD and M-ED are not suitable for weak coupling scenarios: EF-LD exhibits excessively rapid kinetic energy decay, whereas M-ED obviously overestimates random force, leading to an unexpected increase in kinetic energy when $\Gamma$ is small. 
In EF-LD, both the friction $\gamma$ and the random force $\xi$ are treated under the Markovian approximation, so the detailed balance is obeyed even $\Gamma$ is small. However, the Markovian approximation of EF-LD leads to an overly rapid decay of kinetic energy, causing it to deviate substantially from the actual dynamics observed in weak system-bath coupling regimes. While in M-ED, we only use the Markovian random force but do not make the Markovian approximation in the mean force term, thus the detailed balance breaks down.
Only NM-ED demonstrates reasonable agreement with SH, highlighting the effectiveness of our non-Markovian random force in capturing certain characteristics of weakly coupled systems.
As $\Gamma$ increases, the average kinetic energy by all methods except the traditional ED gradually matches. 
In Fig~\ref{fig:eqpo}(a) and (b), we plot the electronic population at the impurity level as a function of time. Notice that the population in the impurity level of the EF-LD is calculated as follows,
\begin{eqnarray}
\begin{aligned}
	N(t) = &\int dxdpP_1(x,p,t) \\ \approx &\int dxdpf(h)A(x,p,t) 
	= <f(h)>.
\end{aligned}
\end{eqnarray}

Notably, the population obtained through ED differs significantly from that obtained through all other methods, rendering its reliability extremely low.
When $\Gamma$ is small, the results obtained with SH and NM-ED exhibit the highest similarity. 
As $\Gamma$ increases to match $\hbar\omega$, 
the results of EF-LD and M-ED in the calculation of population are closer to those of SH and NM-ED.
\begin{figure}[ht]
\centering
\includegraphics[width = 8cm]{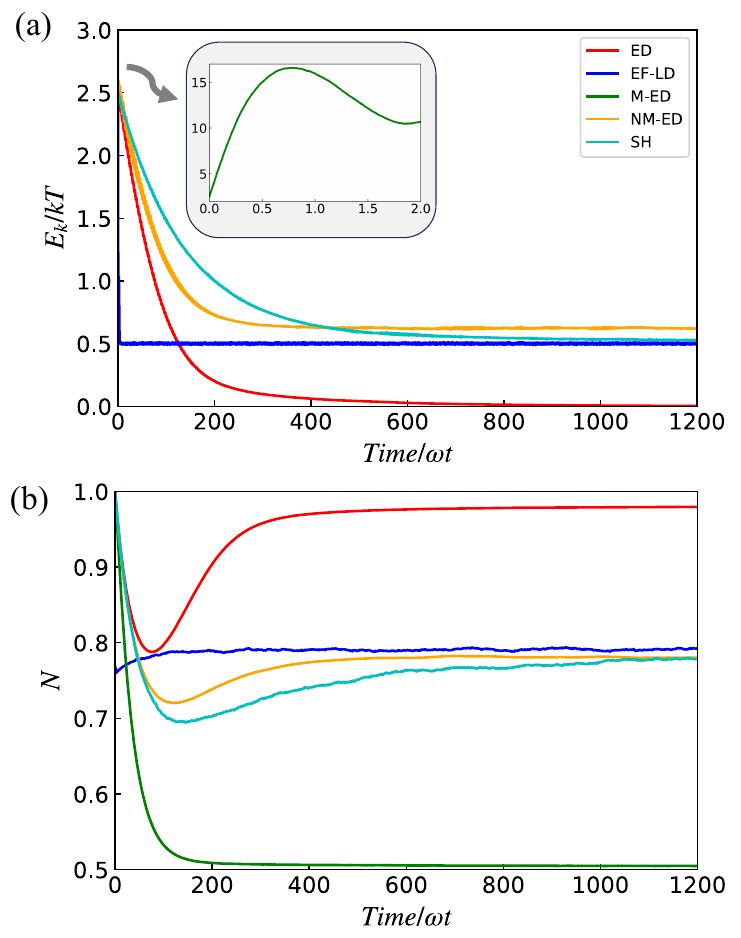}
\caption{Phonon relaxation and electronic population in the impurity level by five methods when the system-bath coupling is extremely small. Parameters: $\hbar\omega = 0.003, kT = 0.05, g = 0.02, \ E_d = g^2/2\hbar\omega$, and $\Gamma = 0.0001$. 
We prepare the initial states satisfying a Boltzmann distribution with a temperature of $5kT$. Due to M-ED (green curve) has an incorrect rapid increase in kinetic energy from the very beginning in (a), its behavior cannot be meaningfully compared with the other methods on the main plot. Therefore, we plot the M-ED curve separately in the inset. The results show that only NM-ED and SH match well.}
\label{fig:small}
\end{figure}

\begin{figure}[ht]
\centering
\includegraphics[width = 8cm]{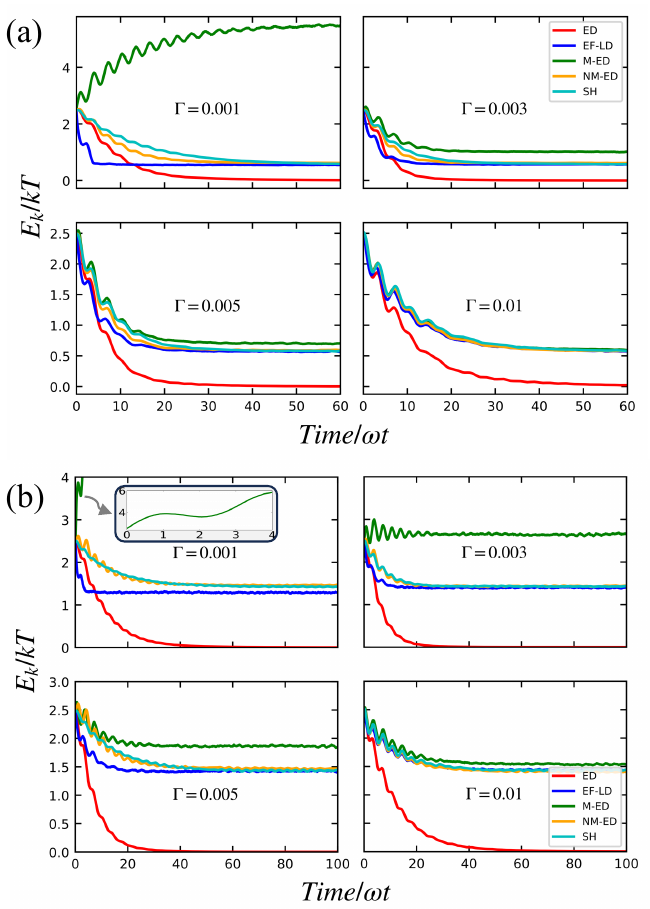}
\caption{Phonon relaxation at different $\Gamma$. Here, we plot the nuclear kinetic energy via five different methods as a function of time. We set $\Gamma_{\text{L}} = \Gamma_{\text{R}} = \Gamma/2$ and $\mu_{\text{L}}=-\mu_{\text{R}}$.
Parameters: $\hbar\omega = 0.003, kT = 0.05, g = 0.02, E_d = g^2/2\hbar\omega$, and $\mu_{\text{L}} = 0.05$ in (a) while $\mu_{\text{L}} = 0.2$ in (b).
We prepare the initial states satisfying a Boltzmann distribution with a temperature of $5kT$. Due to M-ED (green curve) having an incorrect rapid increase in kinetic energy from the very beginning in (b), its behavior cannot be meaningfully compared with the other methods on the main plot. Therefore, we plot the M-ED curve separately in the inset. NM-ED agrees with SH when $\mu$ is large, even though $\Gamma$ is small.} 
\label{fig:neqek}
\end{figure}

\begin{figure}[ht]
\centering
\includegraphics[width = 8cm]{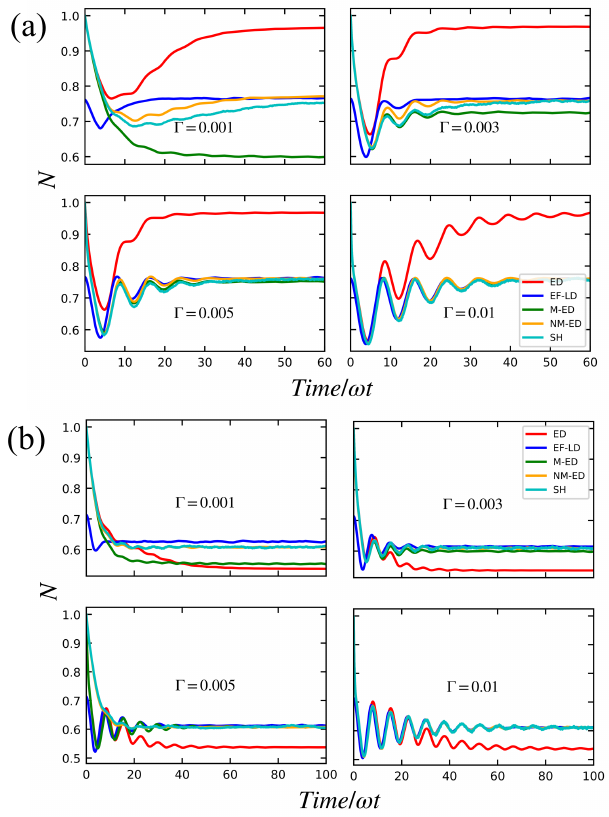}
\caption{Electronic population in the impurity level by five different methods. 
Parameters: $\hbar\omega = 0.003, kT = 0.05, g = 0.02, E_d = g^2/2\hbar\omega$, and $\mu_{\text{L}} = 0.05$ in (a) while $\mu_{\text{L}} = 0.2$ in (b). 
NM-ED agrees with SH when $\mu$ is large even though $\Gamma$ is small.}
\label{fig:neqpo}
\end{figure}
When $\Gamma$ far exceeds $\hbar\omega$, the results obtained from several methods, excluding ED, are highly consistent.
To further investigate the performance of the NM-ED method in extremely small system-bath coupling situations, we test the phonon relaxation and electronic population at $\Gamma = 0.0001$ in terms of five different trajectory-based methods. The results are shown in Fig.~\ref{fig:small}. 
Because of the overestimation of the Markovian random force, the kinetic energy rises rapidly in a short time going beyond the limitations of the y-axis, so that we almost cannot see that curve in Fig~\ref{fig:small}(a), and we have to put the result of M-ED into the inset. Compared with the results at $\Gamma = 0.001$ in Fig.~\ref{fig:eqek}(b) and Fig.~\ref{fig:eqpo}(b), we know that only NM-ED and SH still have a good match in the case of extremely weak coupling, which proves that the non-Markovian random force we constructed is reasonable and has strong applicability. 
The only regret is that the steady-state value of $E_k$ in the long-time limit does not precisely converge to the correct equilibrium solution, where the expected value of average kinetic energy should be $\frac{1}{2}kT$. Therefore, the detailed balance is not perfectly maintained under this extremely weak coupling condition.

\subsection{Case 2: two electronic baths}
\label{sec:case2}
Now, consider the case where a molecule is subjected to two baths with unequal Fermi energies, which we call a non-equilibrium case. 
For such a case, we must use the effective Fermi function $f_{\text{eff}}(h)$ substituting the original one \cite{dou2015surface},
\begin{eqnarray}
    f_{\text{eff}}(h) = \frac{\Gamma_{\text{L}} f_{\text{L}}(h)+\Gamma_{\text{R}} f_{\text{R}}(h)}{\Gamma},
\end{eqnarray}
where $f_{\text{L}}$ and $f_{\text{R}}$ indicate the Fermi functions for the left and right baths, while $\Gamma_{\text{L}}$ and $\Gamma_{\text{R}}$ are the coupling strengths between the system and the bath on both sides, $\Gamma = \Gamma_{\text{L}}+\Gamma_{\text{R}}$.

Similar to the equilibrium case, we run the dynamics using five methods, which are ED, EF-LD, M-ED, NM-ED, and SH, respectively.
The results depicted in Fig. \ref{fig:neqek} and Fig. \ref{fig:neqpo}, which are also compared with those in Fig. \ref{fig:eqek}(b) and Fig. \ref{fig:eqpo}(b), unveil a surprising trend: with the increase in chemical potential $\mu$, the agreement between NM-ED and SH improves notably, even at very low coupling strengths ( $\Gamma = 0.001$). 

\begin{figure}[ht]
\centering
\includegraphics[width = 8cm]{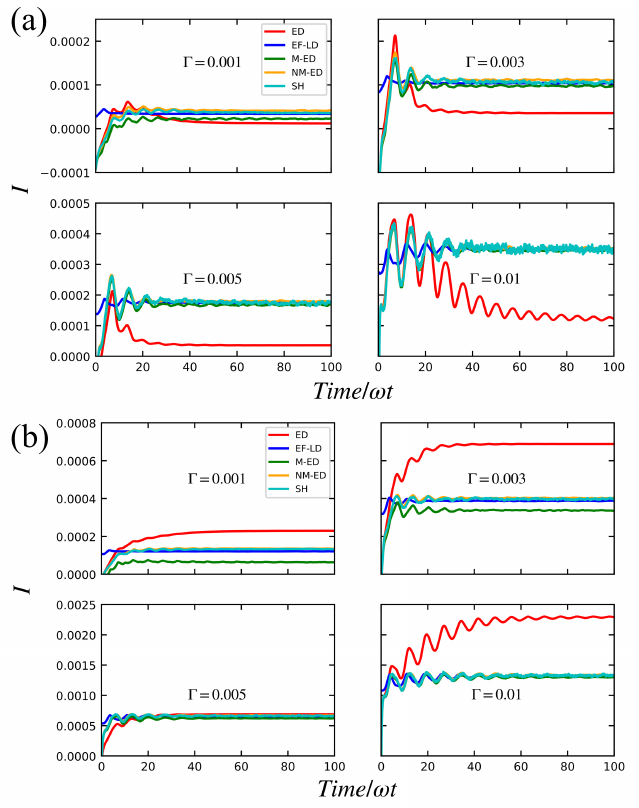}
\caption{The electronic current with time by five different methods.
Parameters: $\hbar\omega = 0.003, kT = 0.05, g = 0.02, E_d = g^2/2\hbar\omega$, and $\mu_{\text{L}} = 0.05$ in (a) while $\mu_{\text{L}} = 0.2$ in (b).}
\label{fig:neqcu}
\end{figure}

In the non-equilibrium scenario, an important observable in quantum transport is the electron current, which can be calculated over time for a biased conduction junction with $f_{\text{L}} \neq f_{\text{R}}$. The electronic current $I$ is given by the integral
\begin{eqnarray}
    I = \int dxdp (\gamma ^{\text{L}} _{0\rightarrow 1}(x)P_0(x,p) - \gamma  ^{\text{L}} _{1\rightarrow 0}(x)P_1(x,p)), 
\end{eqnarray}
where
\begin{eqnarray}
\begin{aligned}
    \gamma ^{\text{L}} _{0\rightarrow 1} =& \frac{\Gamma_{\text{L}}}{h}f_{\text{L}}(\Delta V),\\
    \gamma ^{\text{L}} _{1\rightarrow 0} = &\frac{\Gamma_{\text{L}}}{h}(1-f_{\text{L}}(\Delta V)).
\end{aligned}    
\end{eqnarray}
The numerical results are shown in Fig. \ref{fig:neqcu}. Generally, the electron current increases with the system-bath coupling strength voltage $\mu_{\text{L}}$, which can be understood easily through Ohm's law. 
Interestingly, the electron current also rises with the system-bath coupling strength ($\Gamma$), this is because a higher coupling strength means that electrons within the system are more likely to be scattered or transmitted due to interactions with baths, which typically results in an increase in current.
Notably, when the coupling is weak ($\Gamma = 0.001$), only the electron current obtained by NM-ED and SH exhibit good agreement.

\subsection{Compare analytical and numerical methods}
\label{sec:compare}
In Sec. \ref{sec:compare}, we will compare the dynamics results of NM-ED obtained by the method introduced in Sec. \ref{sec:rf1} and Sec. \ref{sec:rf2}.
Though the two method use same SDE, the most obvious difference between these two methods is the source of $[\overline{D}^S_{\alpha \nu}]_M$.
In the first method we mentioned in Sec. \ref{sec:rf1}, we directly calculate $[\overline{D}^S_{\alpha \nu}]_M$ by Eq. \ref{eq:friction_model} in EF model; while in the second method mentioned in Sec. \ref{sec:rf2} we numerically fit $[\overline{D}^S_{\alpha \nu}]_M$ according to the power spectrum of the correlation function, which is a more general method.

\begin{figure}[ht]
\centering
\includegraphics[width = 8cm]{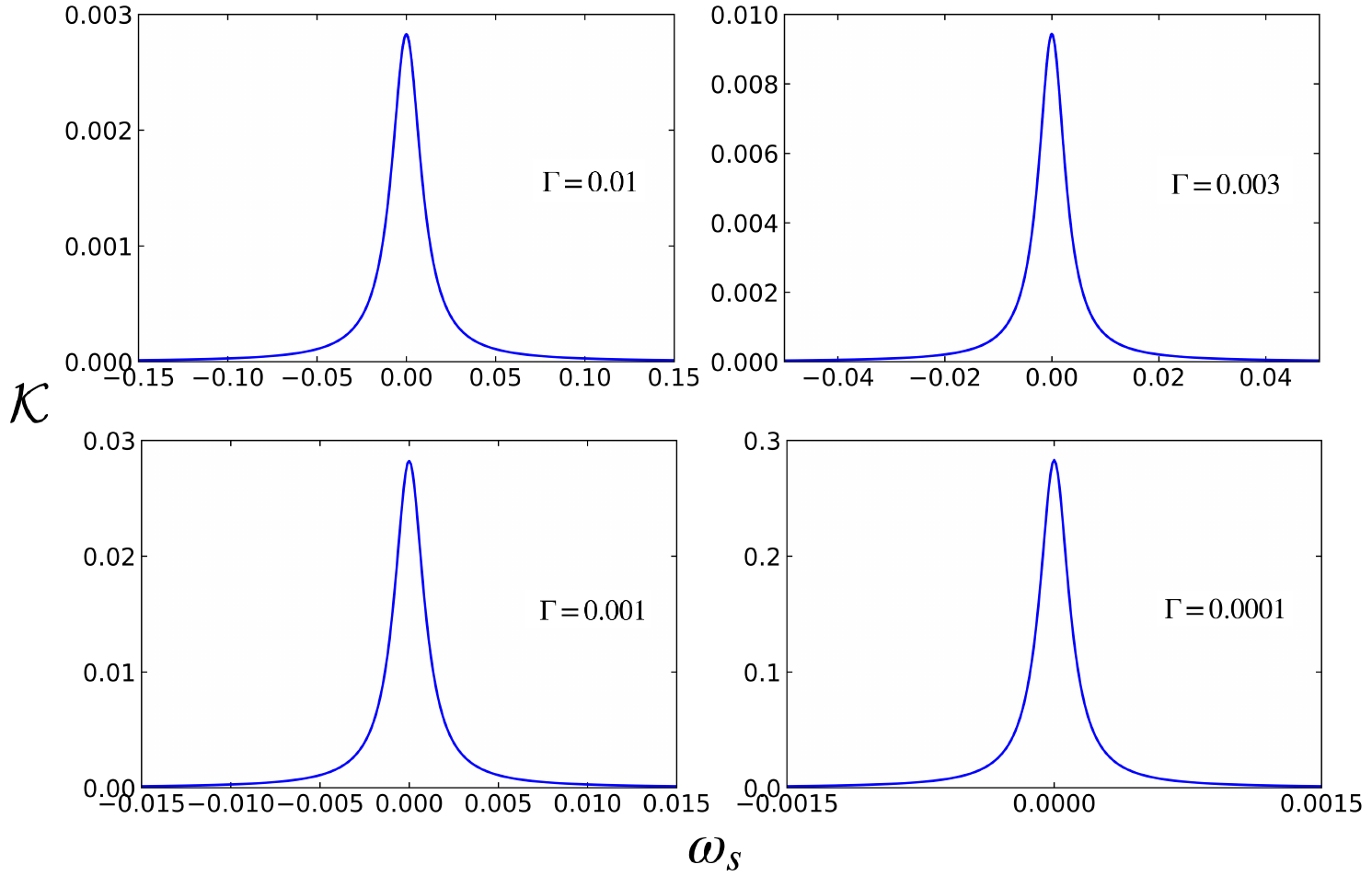}
\caption{ The power spectrum of the correlation functions for RF with different $\Gamma$.
Parameters: $\hbar\omega = 0.003, kT = 0.05, g = 0.02, E_d = g^2/2\hbar\omega$, $\mu_{\text{L}} = 0$, and $x = - \sqrt{2}g/h\omega$.}
\label{fig:spectrum}
\end{figure}

\begin{figure}[ht]
\centering
\includegraphics[width = 8cm]{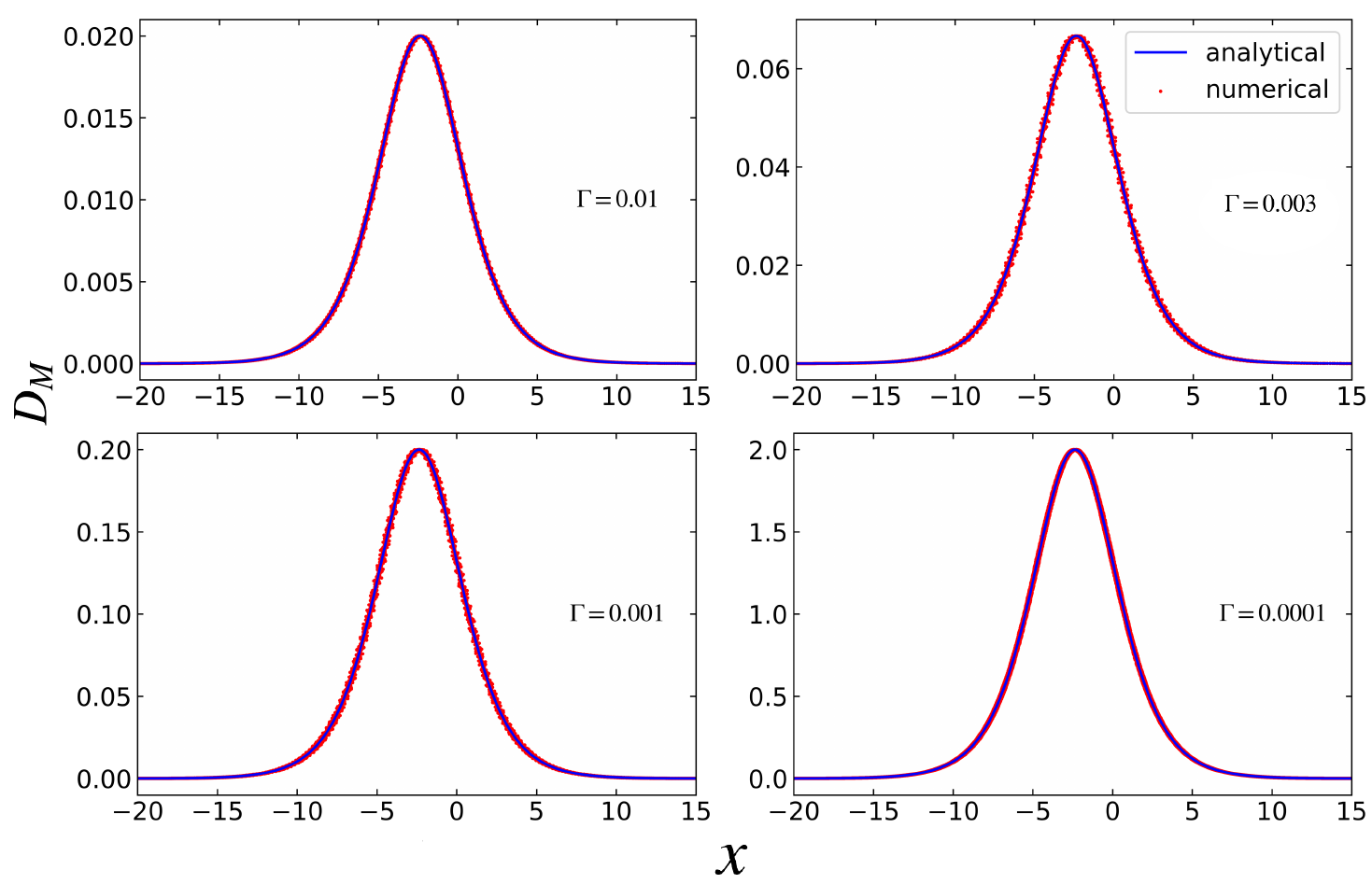}
\caption{ Comparison of $[\overline{D}^S_{\alpha \nu}]_M$ 
 with respect to $x$ calculated by the analytical method 1 mentioned in Sec. \ref{sec:rf1} and fitted by the numerical method 2 mentioned in \ref{sec:rf2}.
Parameters: $\hbar\omega = 0.003$, $kT = 0.05$, $g = 0.02$, $E_d = g^2/2\hbar\omega$, and $\mu_{\text{L}} = 0$. The results obtained by these two methods show a very good match.}
\label{fig:dm}
\end{figure}

We first plot the power spectrum of the correlation function for random force with different set $\Gamma$, which is shown in Fig. \ref{fig:spectrum}. 
Based on the shape of the spectral lines, we can see that they are very obvious Lorentzian spectrum.
Before verifying the dynamics results from the two methods, we first compare the $[\overline{D}^S_{\alpha \nu}]_M$ values obtained by each. 
As shown in Fig. \ref{fig:dm}
, the numerical and analytical $[\overline{D}^S_{\alpha \nu}]_M$ are in perfect agreement. 
In addition, the fitted $\Gamma^{\prime}$ also agrees well with the $\Gamma$ we set.

\begin{figure}[ht]
\centering
\includegraphics[width = 8cm]{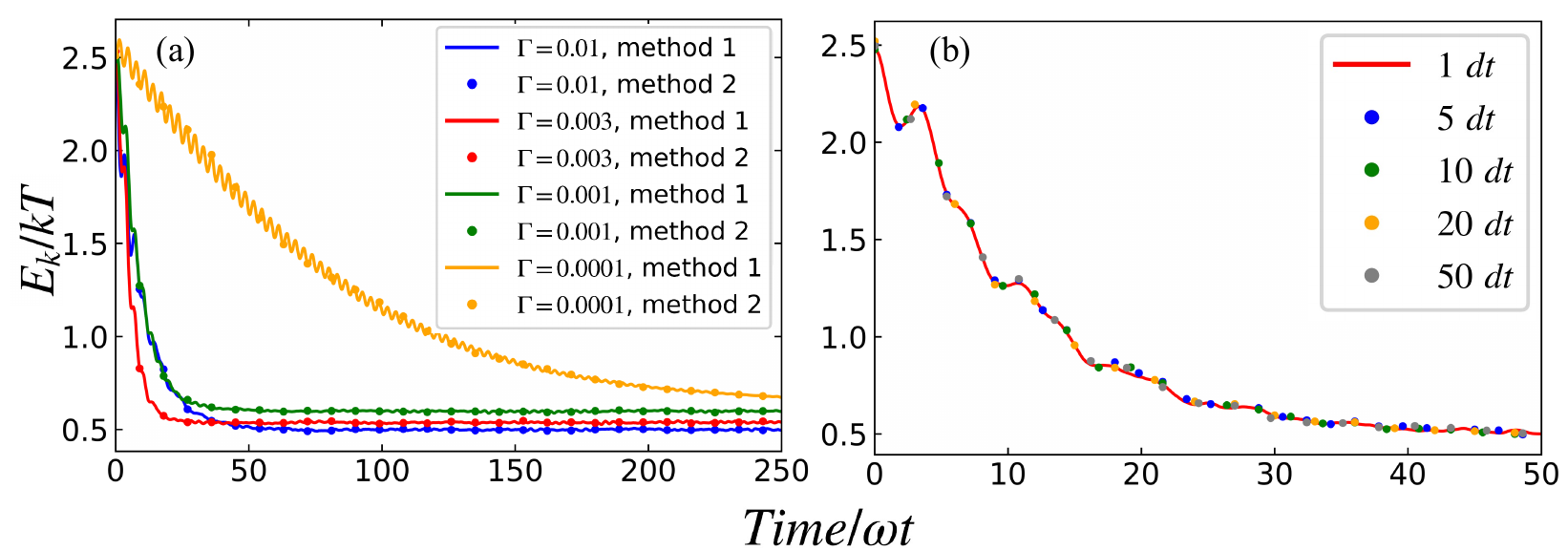}
\caption{ (a) Comparison of phonon relaxation
 with respect to time calculated by the analytical method 1 mentioned in Sec. \ref{sec:rf1} and fitted by the numerical method 2 mentioned in \ref{sec:rf2}. 
In both cases the numerical results of two method match well.
 (b)  Comparison of phonon relaxation
 with respect to time calculated by method 2 with update $[\overline{D}^S_{\alpha \nu}]_M$ in different time step when $\Gamma = 0.01$.
Parameters: $dt = 1$, $\hbar\omega = 0.003$, $kT = 0.05$, $g = 0.02$, $E_d = g^2/2\hbar\omega$, and $\mu_{\text{L}} = 0$. }
\label{fig:dynamics_com}
\end{figure}

We compare the phonon relaxation calculated by two methods, as shown in Fig. \ref{fig:dynamics_com}(a). The results obtained by the two methods have almost no numerical error with different $\Gamma$. 
In addition, considering that updating $[\overline{D}^S_{\alpha \nu}]_M$ at every small time step ($dt = 1$) would incur a huge computational cost, we  update $[\overline{D}^S_{\alpha \nu}]_M$  at intervals in Fig. \ref{fig:dynamics_com}(b).
The results show that even if we update $[\overline{D}^S_{\alpha \nu}]_M$ every 50 time steps, it has almost no effect on the dynamic results, which can greatly reduce our calculation cost.
Note that the update rate of $[\overline{D}^S_{\alpha \nu}]_M$ is very system-dependent. For each new system, we need to retest the convergence of the dynamics at different update rates.

\section{Conclusions}\label{sec:conclusions}
In conclusion, we present a general method to improve ED by incorporating a random force (E$+$R). ED with a Markovian random force (M-ED) proves effective primarily in scenarios with strong system-bath coupling. However, for smaller system-bath couplings, the Markovian random force tends to be overestimated, resulting in implausible outcomes. Encouragingly, our proposed ED approach with a non-Markovian random force (NM-ED) demonstrates robust performance across various coupling strengths in both equilibrium and non-equilibrium conditions. Notably, it achieves remarkable accuracy even under extremely weak coupling conditions. Additionally, under non-equilibrium conditions, the discrepancies between NM-ED and SH results decrease as the chemical potential difference between the two baths increases.
Furthermore, we verify that updating the correlation function of random force $[\overline{D}^S_{\alpha \nu}]_M$ even in large
steps can still maintain the accuracy, significantly reducing the computational cost.

In the current work, we restrict ourselves to Fermionic baths and demonstrate that introducing a random force effectively restores detailed balance in this context. 
While ED is also widely used in systems with Bosonic baths, such as the spin-boson model, we anticipate that adding the appropriate random force can similarly address the lack of detailed balance in such cases. 
Additionally, we believe that our E$+$$\sigma$ method has the potential to be extended to systems with coordinate-dependent coupling and to dynamics involving transitions across multiple basins. 
Furthermore, future studies will also aim to apply the E$+$$\sigma$ framework to more realistic and complex systems to thoroughly evaluate its performance and feasibility in practical applications. 
That being said, how to add the correct random force in these cases is under investigation.

At last, we must admit that the efficiency of our method heavily relies on the convergence of the random force correlation function. While this remains a significant challenge for more complex systems, it is very possible that one could better estimate the correlation function approximately. Despite these challenges, our work provides new perspective of addressing the shortcomings of ED, paving the way for future studies on more complex and realistic systems. This journey is ongoing.

\section{Acknowledgement}
W.D. acknowledges funding from the National Natural Science Foundation of China (Grant Nos. 22273075 and 22361142829) and the Zhejiang Provincial Natural Science Foundation (Grant No. XHD24B0301).



\bibliography{ref.bib}

\end{document}